\documentclass[10pt,conference]{IEEEtran}
\IEEEoverridecommandlockouts
\usepackage{cite}
\usepackage{amsmath,amssymb,amsfonts}
\usepackage{algorithm}
\usepackage{algorithmicx}
\usepackage[noend]{algpseudocode}
\usepackage{graphicx}
\usepackage{textcomp}
\usepackage{xcolor}

\usepackage{url}

\usepackage{booktabs} 
\usepackage{tabularx}
\usepackage{caption}
\usepackage{subcaption}
\usepackage{amsmath}
\usepackage{balance} 
\usepackage{amssymb}
\usepackage{xspace}
\usepackage[T1]{fontenc}
\usepackage[scaled=0.81]{beramono}
\usepackage{balance}
\let\labelindent\relax
\usepackage{enumitem}  
\usepackage{listings}
\usepackage{url}
\usepackage{multirow}
\usepackage{array}
\usepackage{color}
\usepackage{float}
\usepackage[misc]{ifsym}
\usepackage{booktabs}
\usepackage{pdfpages}

\newcommand{\eg}{{\emph{e.g.}} }

\newcommand{\tool}{TrustCross\xspace}

\def\BibTeX{{\rm B\kern-.05em{\sc i\kern-.025em b}\kern-.08em
		T\kern-.1667em\lower.7ex\hbox{E}\kern-.125emX}}
\begin{document}
	
\title{\tool: Enabling Confidential Interoperability across Blockchains Using Trusted Hardware}

\author{
	\IEEEauthorblockN{
		Ying Lan\IEEEauthorrefmark{1}, 
		Jianbo Gao\IEEEauthorrefmark{2}\IEEEauthorrefmark{3},
		Ke Wang\IEEEauthorrefmark{2}\IEEEauthorrefmark{3},
		Jiashuo Zhang\IEEEauthorrefmark{2}\IEEEauthorrefmark{3},
		Zhenhao Wu\IEEEauthorrefmark{2}\IEEEauthorrefmark{3},
		Yuesheng Zhu\IEEEauthorrefmark{1},
		Zhong Chen\IEEEauthorrefmark{2}\IEEEauthorrefmark{3}\IEEEauthorrefmark{4}
	}
	\IEEEauthorblockA{\IEEEauthorrefmark{1}\textit{School of Electrical and Computer Engineering}, \textit{Peking University}, Shenzhen, China}
	\IEEEauthorblockA{\IEEEauthorrefmark{2}\textit{Department of Computer Science and Technology, EECS}, \textit{Peking University}, Beijing, China}
	\IEEEauthorblockA{\IEEEauthorrefmark{3}\textit{Key Laboratory of High Confidence Software Technologies (Peking University), MoE}, Beijing, China}
	\IEEEauthorblockA{\IEEEauthorrefmark{4}Corresponding Author}  
	\{ly3996, gaojianbo, wangk, zhangjiashuo, zhenhaowu, zhuys, zhongchen\}@pku.edu.cn

}

\maketitle

\begin{abstract}
With the rapid development of blockchain technology, 
different types of blockchains are adopted and interoperability across blockchains has received widespread attention.
There have been many cross-chain solutions proposed in recent years, including notary scheme, sidechain, and relay chain. 
However, most of the existing platforms do not take confidentiality into account, although privacy has become an important concern for blockchain.
In this paper, we present \tool, a privacy-preserving cross-chain platform to enable confidential interoperability across blockchains.
The key insight behind \tool is to encrypt cross-chain communication data on the relay chain with the assistance of trusted execution environment and employ fine-grained access control to protect user privacy.
Our experimental results show that \tool achieves reasonable latency and high scalability on the contract calls across heterogeneous blockchains.
\end{abstract}


\section{Introduction}
\label{sec:intro}


As the underlying technology of cryptocurrencies and trusted decentralized applications, blockchain has developed rapidly in recent years and has become one of the most attention-grabbing technologies in both academia and industry. 
The number of blockchain-based applications has increased massively in different fields, such as finance, internet of things, cloud storage, digital rights and assert management\cite{wust2018you}\cite{gao2019towards}.
With the gradual in-depth research of blockchain technology, more and more blockchain platforms appear in various scenarios.
The isolation between different blockchains inevitably results in the islanding effect and interoperability across heterogeneous blockchains has received widespread attention since.

To enable cross-chain interoperability, different solutions have been proposed. 
Notary scheme\cite{qasse2019inter} introduces a trusted intermediary that facilitates information exchange and mediates between blockchains, \eg Ripple's ILP protocol\cite{thomas2015protocol}. 
Another approach is side chain\cite{qasse2019inter}, which is a designation for a blockchain ledger that runs in parallel to a primary blockchain. 
Entries from the primary blockchain (where said entries typically represent digital assets) can be linked to and from the sidechain. 
This allows the sidechain to otherwise operate independently of the primary blockchain. 
Back et al.\cite{back2014enabling} proposed a pegged sidechain to enable Bitcoin and other ledger assets to be transferred between multiple blockchains. 
While notary scheme and sidechain is mostly used in cryptocurrency exchange, relay chain is the common solution in industry.
Relay chain was first proposed in the chain fibers scheme\cite{chain-fibers}, which proposes a concept of cross-chain data exchange between a single relay chain and multiple isomorphic chains, using the characteristics of decoherence and transaction Lantecy to coordinate the transactions of multiple parts of the system. 
Afterwards, various cross-chain projects including Polkadot\cite{wood2016polkadot}  and Cosmos\cite{cosmos2021} have been improved on this basis, enriching the original cross-chain architecture, enabling it to exchange data and assets in a variety of heterogeneous chains.

Although the aforementioned approaches meet the needs of cross-chain interaction to a certain extent, they do not solve an important problem: data privacy. In some specific business or application scenarios, data privacy leakage will be a serious problem. For example, in applications such as health tracking, disease treatment, scientific research, etc.\cite{kuo2017blockchain}, cross-chain sharing of biomedical data is usually involved. These data are closely related to personal life information, and even contain information such as genes and proteins. If information is leaked, it will cause big social problems, such as genetic discrimination. 


In this paper, we present \tool, a privacy-preserving cross-chain platform to enable confidential interoperability across blockchains based on trusted execution environment (TEE). 
TEE is a tamper-resistant processing environment that provides runtime isolation and guarantees the integrity of the runtime states. The execution code and data in TEE cannot be accessed by outsiders. We utilize these properties of TEE and combine the relay chain to achieve confidential cross-chain interoperability. 

The main contributions of this paper can be summarized as follows:
\begin{itemize}
	\item \textbf{\tool Architecture}. We propose \tool, a novel cross-chain architecture that enhances confidentiality with TEE. 
	\item \textbf{Cross-chain Interoperability Protocol}. We design a cross-chain protocol to enable a blockchain client to execute transactions on different blockchain systems. Because different blockchains have different transaction structures, so it's necessary to have a unified standard format to facilitate cross-chain interoperability
	\item \textbf{Access Control Mechanism}. We design a fine-grained access control mechanism to prevent data from being abused. Access control policies can prevent malicious users from accessing data on the blockchain (such as smart contract functions) through brute force enumeration.
	\item \textbf{Security Key Exchange}. In order to ensure the confidentiality of cross-chain data transmission between the blockchain and the relay chain, we have design a secure key exchange algorithm.
\end{itemize}

The rest of this paper is organized as follows.
Section~\ref{backgroud} provides a brief background on blockchain and TEE. 
Section~\ref{scheme} presents the overview of \tool and the details of key management and workflow are described in Section~\ref{keymanagement}. 
We analyze the security attributes provided by \tool in Section~\ref{analysis} and evaluate the platform in Section~\ref{evaluation}
In section \ref{relatedwork}, we introduce and discuss the related work. Finally, we have a conclusion and future work discussion in section \ref{conclusion}.

\section{Background}
\label{backgroud}
In this section, we briefly introduced the concepts of blockchain, TEE and Intel SGX, and explained some of their properties used in our proposal. 

\subsection{Blockchain}
Blockchain\cite{2009bitcoin} is a distributed ledger, which has the characteristics of decentralization, tamper-proof, traceability, transparency. These characteristics are guaranteed by distributed system structure, consensus algorithm and cryptographic methods. Different blockchains are different in system architecture, consensu algorithms, and data structures. But they all have a common chain structure.
Each block in a blockchain contains the hash of the previous block, the hash of the block, the timestamp, the transactions and other information. 
Due to the collision avoidance feature of the hash function, it is difficult to modify the previously confirmed block. This iterative process confirms the integrity of the previous block and goes back to original genesis block. 

Blockchain can be divided into permissioned blockchain and permissionless blockchain. The permissionless blockchain network, such as Ethereum\cite{wood2014ethereum}, can be participated by anyone, while the permissioned blockchain network, such as Fabric\cite{cachin2016architecture}, can only be participated by certain people. At present, many consensus algorithms have been proposed. The more mainstream ones are PoW\cite{2009bitcoin}, PoS\cite{king2012ppcoin}, DPoS\cite{larimer2017dpos}, PBFT\cite{sukhwani2017performance}, etc.

\subsection{TEE and Intel SGX}
Trusted Execution Environment (TEE) is a tamper-resistant processing environment that runs on a separation kernel. It guarantees code and data loaded inside to be protected with respect to confidentiality and integrity\cite{sabt2015trusted}. Modern TEE environment, the most famous ones are ARM TrustZone\cite{alves2004trustzone,winter2008trusted} and Intel SGX\cite{costan2016intel,mckeen2016intel}. In our proposal, Intel SGX is used rather than Trustzone, because Intel SGX provides a key feature unavailable in TrustZone, called attestation, which can prove its trustworthiness to third-parties.

Intel Software Guard Extensions (SGX) is a set of security-related instruction codes that are built into some modern Intel central processing units (CPUs)\cite{intelsgx}. The key abilities provided by Intel SGX are runtime isolation, sealing and attestation.

\textbf{Runtime isolation}. The regions of memory where private code and data run on is called Enclaves. With protective mechanisms enforced in the processor, enclaves are protected and unable to be accessed by any process outside the enclave itself, including processes running at higher privilege levels. All runtime enclave memory is encrypted, and the state cannot be observed. The code and data of an enclave are stored in a protected memory area called Enclave Page Cache (EPC) that resides in Processor Reserved Memory (PRM)\cite{mckeen2013innovative}. 

\textbf{Sealing}. Sealing\cite{intelsgx-sealing} is the process of encrypting enclave secrets for persistent storage to disk. Encryption is performed using a private Seal Key that is unique to that particular platform and enclave, and is unknown to any other entity. There are two identities associated with an enclave. One is the Enclave Identity, another is the Signing Identity. When selecting different identity, sealing can have different functions. Using the Enclave Identity will restrict access to the sealed data only to instances of that enclave, while using the Signing Identity to sealing can share sensitive data via a sealed data between multiple enclaves for a given application and/or enclaves of different applications produced by the same development firm, because Signing Identity is provided by an authority and will be the same for all enclaves signed with the same authority.

\textbf{Attestation}. Attestation\cite{mckeen2013innovative,intelsgx,bao2020blockchain} is the process of verifying that enclave code has been properly initialized. SGX provides two types of attestation: intra-attestaion and remote attestation. In intra-attestation, SGX provides the instruction that helps an enclave to attest to another enclave on the same platform directly. Specically, the enclave will generate the certification called REPORT, containing the measurement information. In remote attestation, SGX provides instruction to enable an enclave to prove to the remote platform that private code and data are loaded in the enclave correctly and the enclave is running in the SGX-enabled platform. Specifically, the generated certification is called QUOTE.

%

\section{Overview of \tool}
\label{scheme}
\subsection{Hierarchical Architecture}
\begin{figure}[h]
	\centering 
	\includegraphics[width=0.45\textwidth]{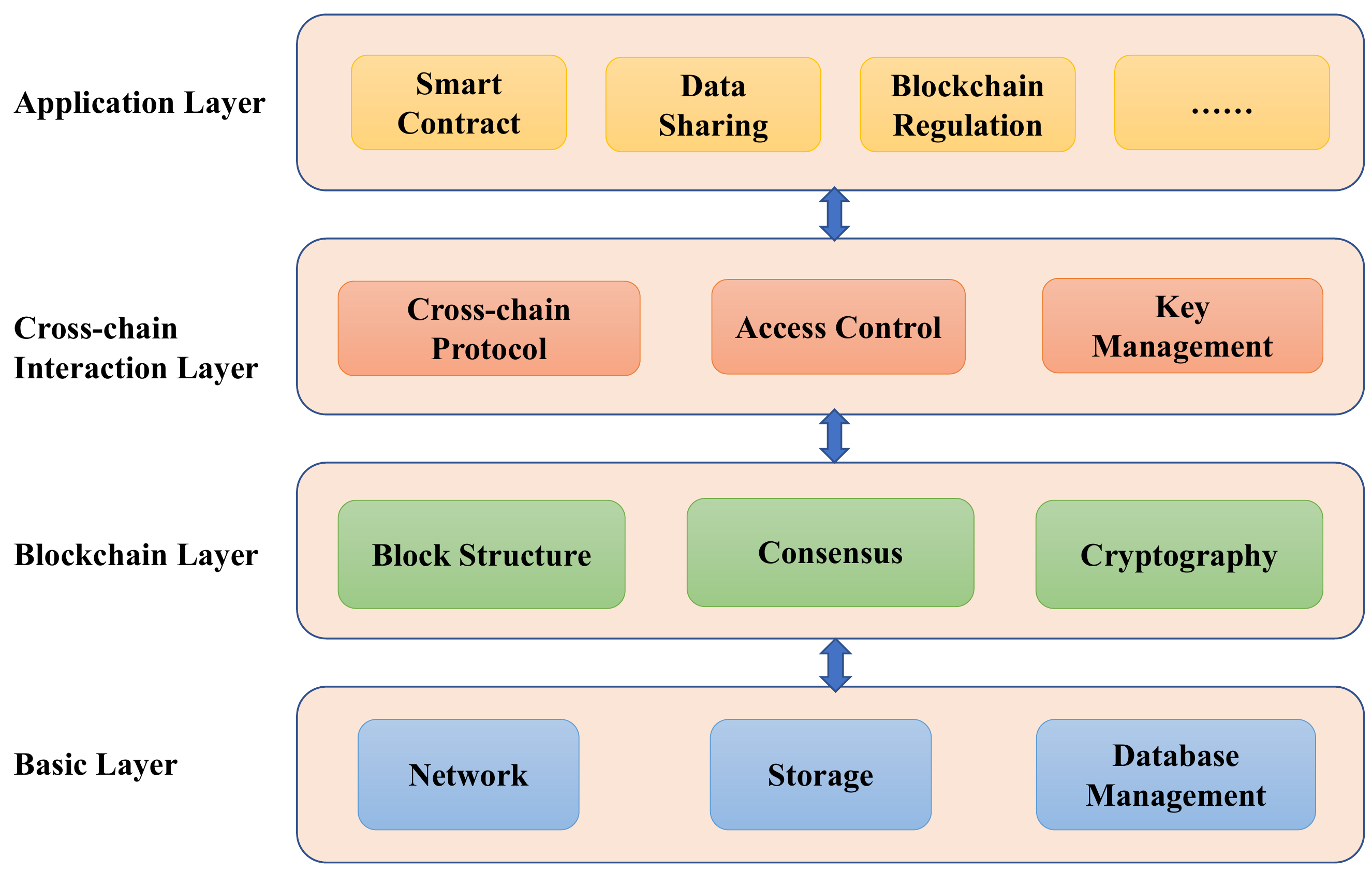} 
	\caption{Hierarchical Architecture}
	\label{Hierarchical Architecture}
\end{figure}

In the existing schemes, an important but not well-solved problem is privacy protection. Thus we propose a hierarchical architecture\cite{2018A} of cross-chain system which shows in Fig.~\ref{Hierarchical Architecture}. 

As it shows in Fig.~\ref{Hierarchical Architecture}, there are four layers: basic layer, blockchain layer, cross-chain interaction layer and application layer. Details regarding to the four layers are given below.

\subsubsection{Basic layer}
The basic layer consists of three parts: network, storage and database management. This layer  is foundation of blockchain layer, providing some basic functions, such as network broadcasting, data storage and access, etc.

\subsubsection{Blockchain Layer}
Various blockchains have been proposed, with different architectures and technical implementations. But there are still some things in common among all blockchains, we briefly summarize it into three modules: block structure, consensus, cryptography. Block is the unit of the ledger. Consensus is an important part of a distributed system. cryptography is the critical technology to ensure block data that is not tampered with.

\subsubsection{Cross-chain Interaction Layer}
Cross-chain Interaction Layer is the main content of our paper. This layer consists of cross-chain protocol, access control, key management.
\begin{itemize}
	\item Cross-chain protocol: In order to achieve cross-chain interoperability between blockchains with different underlying implementations, it is necessary to propose a unified rule to route and execute cross-chain transactions.
	\item Access control: To avoid misuse of important private data, access rights management is required to ensure that the data is used within a legal scope.
	\item Key management: In the process of cross-chain transaction routing, in order to prevent data from being illegal interception, it needs to be encrypted. Therefore, how to securely generate, share and store the key is an important issue.
\end{itemize}

\subsubsection{Application Layer}
With cross-chain interactive services, we can implement smart contract cross-chain operations, data sharing, blockchain regulation and other applications.
\subsection{System Architercture}
\begin{figure}[h]
	\centering 
	\includegraphics[width=0.48\textwidth]{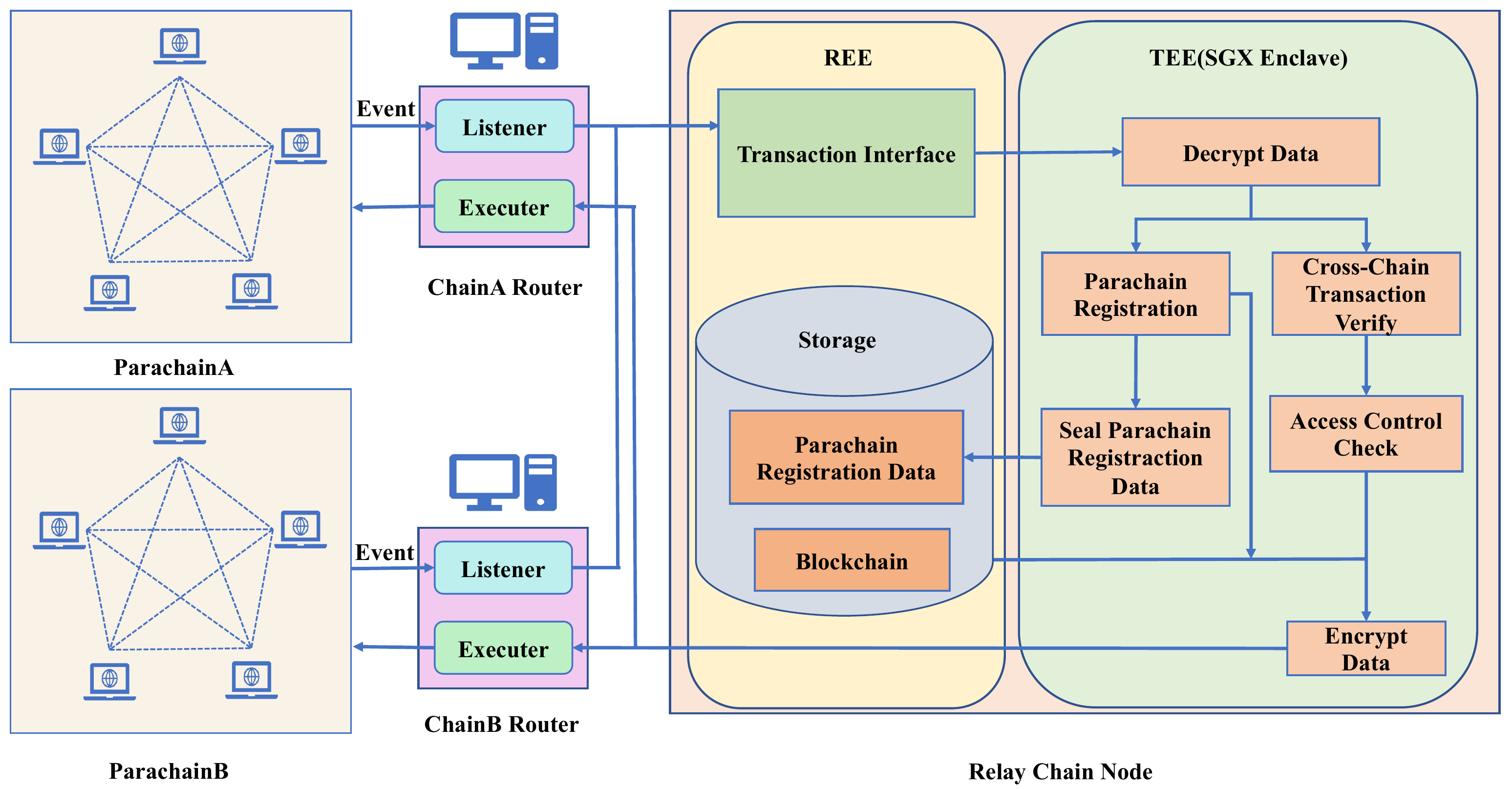} 
	\caption{Architecture overview }
	\label{architecture}
\end{figure}

To enhance the confidentiality of cross-chain, we present \tool, a TEE-based cross-chain scheme. Fig.~\ref{architecture} depicts the architercture of \tool. As it shows, there are three types of roles in this scheme: parachains, routers and relay chain. 

\textbf{Parachains} are those blockchains that are isolated from each other and only interact through relay chain. In our scheme, an user of parachain can start a cross-chain transaction through emitting an event in parachain, which is listened by router.

\textbf{Routers} are the key roles to connect parachains and relay chain. A router is the corresponding blockchain's full node. In our scheme, we consider it an authority. Router is responsible for listening cross-chain events from parachain, and package it into a cross-chain transaction with our cross-chain interoperability protocol and spread it to the relay chain nodes. In addition, the router also undertakes the role of executing cross-chain transactions. Specifically, when receiving a cross-chain transaction from relay chain, router executes it and sends the execution result to relay chain nodes after the transaction is confirmed. Before starting a cross-chain bussiness, router's certificate must be registered with the relay chain.


\textbf{Relay chain}: the main function of the relay chain is to manage parachains, perform trusted verification and access control of cross-chain transactions, record the status of each parachain cross-chain transaction, and forward cross-chain transactions. In our scheme, the relay chain runs the PoA(Proof of Authority) consensus algorithm\cite{poa}. Each relay chain node is equipped with TEE-enabled CPU. As the Fig.~\ref{architecture} shows, a relay chain node can split to REE (Rich Execution Environment) part and TEE (Trust Execution Environment) part. REE has a transaction interface responsible for receiving transactions sent by router, and stores some encrypted data on the disk, such as parachain registration data, blockchain. The encryption and decryption operations of confidential data, the legality verification of cross-chain transactions, and the verification of access rights are performed in the TEE.

\subsection{Cross-chain Interoperability Protocol}
\label{protocol}
\begin{figure}[h]
	\centering 
	\includegraphics[width=0.45\textwidth]{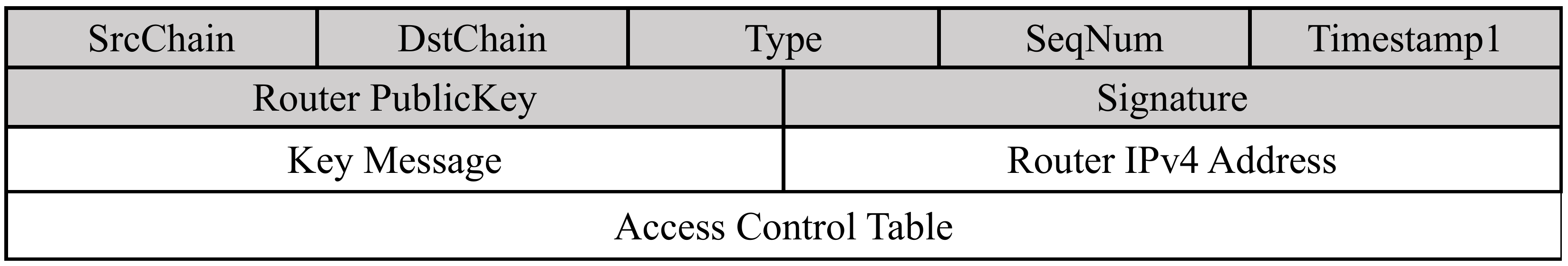} 
	\caption{Registration Transaction}
	\label{register_transaction}
\end{figure}

To unify the format of cross-chain interoperability between different blockchains, we designed a cross-chain interoperability protocol(CCIP). The protocol includes the protocol header and payload content, which we will describe in detail below. 

\subsubsection{Protocol Header} 
contains routing information.

Specifically, the CCIP protocol header contains the source parachain and destination parachain information used for routing, as well as the identity information of the router that packaged the transaction, and so on. The specific fields and meanings are as follows:

\begin{itemize}
	\item SrcChain:  it represents the starter parachain of transactions. Each parachain has a unique ID in our sheme.
	\item DstChain: as opposed to SrcChain, DstChain is the receiver of transactions. 
	\item Type:  distinguish the different functions of the transaction, such as registration, data request, response. 
	\item SeqNum: the unique identification number of the transaction.
	\item Timestamp1: the time when router packages this routing transaction
	\item Router PublicKey: Router's public key which can both identify his identity and check the signature of transaction.
	\item Signature: the transaction will be signed by router to prevent from being tampered with.
	\item Payload: details of the transaction. According to the type field, there are different contents. We will describe in detail in the following paragraphs.
\end{itemize}

\subsubsection{Payload}
specific information of the transaction.

As mentioned earlier, different types will point to different payload information. In general, it can be divided into registration transactions and cross-chain transactions. 

The format of the registration transaction is shown in the Fig.~\ref{register_transaction}. In this case, the DstChain is the relay chain ID. In addition to the protocol header, the fields and meanings of the payload content are as follows:

\begin{itemize}
	\item Key Message: the key message sended to relay chain nodes will compute a same communication key between router and relay chain nodes. This will be described in detail in section \ref{keymanagement}.
	\item Router IPv4 Address: router's IPv4 address is used for the relay chain to forward cross-chain transactions.
	\item Access Control Table: router will register the corresponding parachain's access control table. The design of the access control table will be in section \ref{access_control}. 
\end{itemize}

The format of the registration transaction is shown in the Fig.~\ref{cross-chain-format}. The Payload field will contain the following content:
\begin{figure}[h]
	\centering 
	\includegraphics[width=0.45\textwidth]{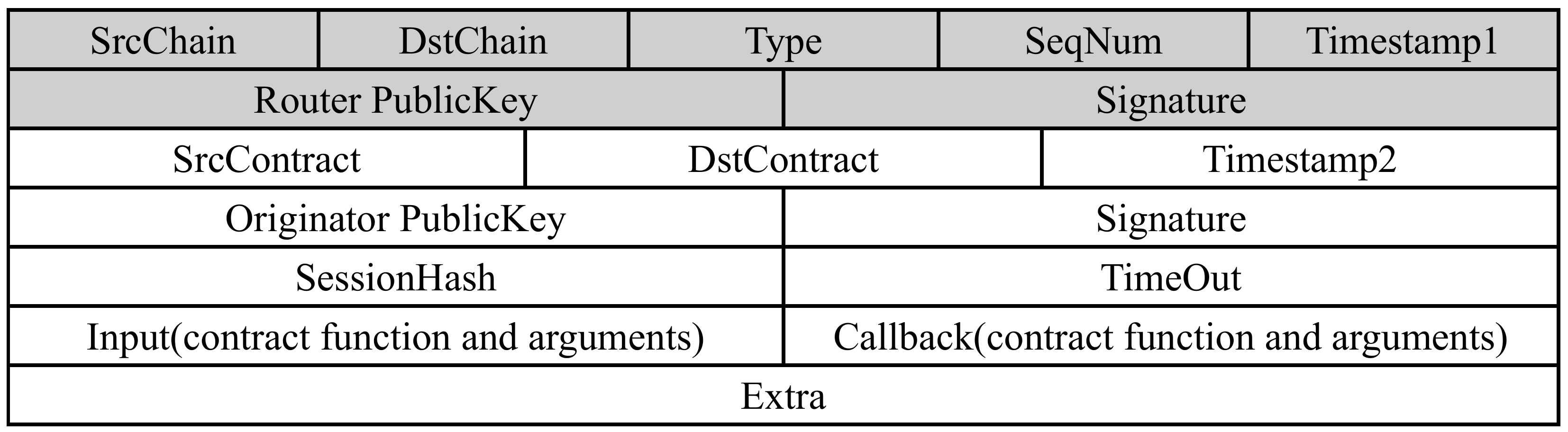} 
	\caption{Cross-chain Transaction}
	\label{cross-chain-format}
\end{figure}

\begin{itemize}
	\item SrcContract: source contract ID from SrcChain.
	\item DstContract: destination contract ID in DstChain. 
	\item Timestamp2: the time of parachain emit the cross-chain event. Used to calculate the valid time of this cross-chain transaction.
	\item Originator PublicKey: the public key of the parachain user who originally initiated the cross-chain transaction. Used for fine-grained user access control and data integrity check.
	\item Signature: originator's signature on cross-chain event data.
	\item SessionHash: we define a complete cross-chain interoperation as a session. If it is the beginning of a session (type=1), this field is 0. If it is a response to the previous transaction (type=2), this field is the hash of the transaction that it responded to.
	\item TimeOut: the available duration of this transaction.
	\item Input: functions and parameters of the DstContract.
	\item Callback: functions and parameters of the SrcContract. When having fetched data from DstContract, need to respond to data to SrcContract through this function interface. If no callback is required, this field can be empty.
	\item Extra: reserved field.
\end{itemize}

The router obtains this information through events published by the parachain. Furthermore, in cross-chain transactions, it can also be divided into data request transactions and response transactions according to the Type field, but they share the same format.

\subsection{Access Control}
\label{access_control}
\begin{figure}[h]
	\centering 
	\includegraphics[width=0.48\textwidth]{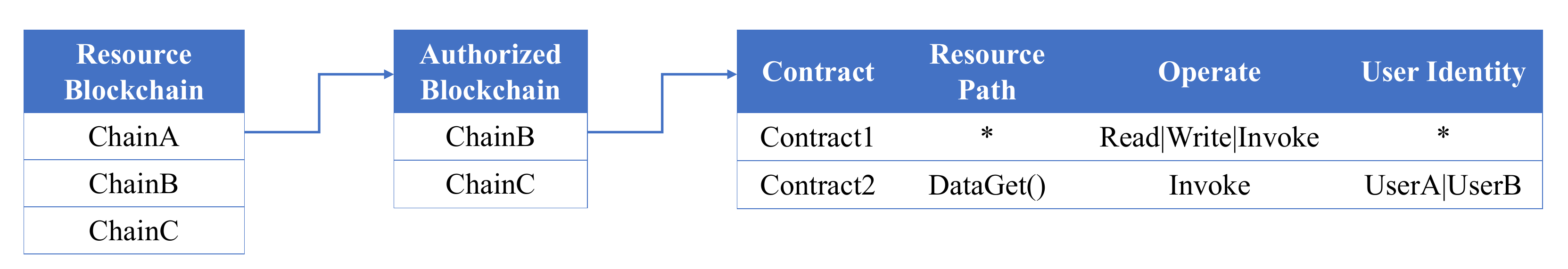} 
	\caption{Access Control List}
	\label{accesscontrol}
\end{figure}

To ensure that legitimate users access protected resources, prevent illegal users from acquiring protected resources, or prevent legitimate users from unauthorized access to protected resources, we propose a fine-grained access control strategy. This strategy subdivides the subject of access control into specific accounts of each parachain rather than the entire parachain. 

An example of access control list is given in Fig.~\ref{accesscontrol}. The meaning of each field is as follows:
\begin{itemize}
	\item Resource Blockchain: the parachain that provides resources, in other words, is the DstChain of cross-chain transactions.
	\item Authorized Blockchain: the parachain that is authorized to access resources, in other words, is the SrcChain of cross-chain transactions.
	\item Contract: the resource owner.
	\item Resource Path: the path of the resource in the contract, such as a function. This field supports wildcards.
	\item Operate: allowed operations on resources, such as read, write or invoke. This field supports wildcards.
	\item User Identity: authorized users, as shown in the Fig.~\ref{accesscontrol}, we use "|" as a separator. This field supports wildcards.
\end{itemize}

\section{Key Management and Workflow Details}
\label{keymanagement}
\subsection{Key Management}
Utilizing the access control mechanism and TEE technology, we enhance the confidentiality guarantee of cross-chain data. But this is still not perfect: those data are still vulnerable while in transit. In the absence of trusted input/output on the relay chain nodes' platform, input may be leaked to malware. Therefore, we need a reliable key generation mechanism to generate the communication key between the router and the relay chain nodes, to ensure that the cross-chain data is encrypted before entering the TEE.

The TEE we use in our scheme is  Intel SGX (Software Gaurd Extensions) \cite{koblitz1987elliptic, costan2016intel}, which provided remote attestation. A platform's enclave can attest to a remote entity that it is trusted, and establish an authenticated communication channel with that entity. As part of attestation, the encalve proves the following:
\begin{itemize}
	\item Its identity.
	\item That it has noe been tampered with.
	\item That it is running on a genuine platform with Intel SGX enabled.
	\item That it is running at the latest security level,  also referred to as the Trusted Computing Base (TCB) level.
\end{itemize}

Utilize the properties listed above, we can use SGX remote attestation to help to realize secure communication key generation and sharing. Our key generation scheme contains two phases. The first phase is the initialization of the relay chain, that is, the nodes of the relay chain generate a common private key through Feldman verifiable secret sharing algorithm\cite{feldman1987practical}. In the second phase, router and relay chain nodes generate a communication key through the ECDH key exchange algorithm\cite{joux2000one}.

\begin{algorithm}
	\caption{Communication Key Generation}
	\label{algorithm2}
	\begin{algorithmic}[1]
		\Require $\mathcal{V}$: validator, $\mathcal{R}$: router, $G$: a base point , $n$: the number of relay chain nodes
		
		\Function{Relay\_nodes\_initial}{ }
		\State $s \gets$ $\mathcal{V}$ generates a random number
		\State $\mathcal{V}$ execute SECRET\_SHARING($s$)
		\For{$i=0 \to n-1$ and $i \neq \mathcal{V}$}
		\State SECRET\_RECOVER()
		\EndFor	
		\EndFunction

		\State 
		\Function{Key\_Exchange}{ }
		\State $\mathcal{R}$ request to start key exchange
		\State $\mathcal{V}$ send ($\mathcal{A}, \sigma$) to $\mathcal{R}$ \Comment{\textcolor{gray}{$\mathcal{A}= s\cdot G$}}
		\State $\mathcal{R}$ verifies remote attestation $\sigma$
		\State $b \gets$ $\mathcal{R}$ generates a random number
		\State $\mathcal{R}$ send $\mathcal{B}$ to $\mathcal{V}$ \Comment{\textcolor{gray}{$\mathcal{B}=b \cdot G$}}
		\State $\mathcal{R}$  gets $key \gets b \cdot \mathcal{A}$
		\State $\mathcal{V}$  gets $key \gets s \cdot \mathcal{B}$
		\EndFunction
		
	\end{algorithmic}
\end{algorithm}

\noindent\textbf{Phase 1}: Relay chain nodes initialize.
\begin{itemize}
	\item[1.] System parameters: Selects two large prime numbers $p$ and $q$, where $q|(p-1)$. $G_q$ is the $q$ order multiplicative subgroups of $Z_{p}^{*}$. Randomly select $g \in G_q$. Let $t = \lceil \frac{2n}{3} \rceil$, where $n$ is number of the relay chain nodes.
	
	\item[2.] Before the relay chain system starts, one of the relay chain nodes generates a 256 bits random number $s$. In our scheme, this node is the validator $\mathcal{V}$ in the PoA\cite{poa} consensus algorithm. 
	
	\item[3.] $\mathcal{V}$ computes a random polynomial 
		\begin{equation}
			f(x)=\prod\limits_{j=0}^{t-1}a_{j}x^{j} \ mod\ q
		\end{equation}
	where $a_0=s \in Z_q, \ a_j \in Z_q (j=1,...,t-1) $.
	
	\item[4.] $\mathcal{V}$ computes shares:
		\begin{equation}
			s_i = f(i) = \prod\limits_{j=0}^{t-1}a_{j}i^{j}\ (i=1,...,n)
		\end{equation}
	 
	\item[5.] For each node $i$ in relay chain, $\mathcal{V}$ establishes an authenticated communication channel with node $i$ through remote attestation, and sends share $s_i$ to node $i$ through the authenticated communication channel.
	
	\item[6.] $\mathcal{V}$ computes commitments: 
		\begin{equation}
			E_j=g^{a_j} \ mod\ p\ (j=0,...,t-1)
		\end{equation}
	 and broadcast those commitsments.
	\item[7.] When node $i$ receives the broadcast commitments and share $s_i$, the following fomula can be calculated to verify the correctness of the share: 
		\begin{equation}
			g^{s_i}=\prod\limits_{j=0}^{t-1}(E_j)^{i^j}\ mod\ p 
		\end{equation}
	
	\item[8.] After the verification is passed, the node $i$ can send its share to neighbors through an authenticated communication channel. If receiving the shares of any t nodes, node $i$ can use Lagrangian interpolation to calculate secret private key $s$:
		\begin{equation}
			s=\sum\limits_{k \in B}(S_k \cdot \prod\limits_{j \neq k}^{t} \frac{j}{j-k} )mod\ q
		\end{equation}
	where $B\subseteq N=\{1,2,...,n\}$, set $B$ is the is the collection of nodes who have sent shares. And node $i$ can verify the secret private key through:
		\begin{equation}
			E_0 \equiv g^s\ mod\ p
		\end{equation}
\end{itemize}

All the above calculations occur in the enclave of the nodes platform. Now we have completed the initialization of the system, and there is a common secret private key $s$ in the enclave of each node.

%
%
%
%

\noindent\textbf{Phase 2}: Key exchange between router and relay chain nodes.
\begin{itemize}
	\item[1.] System parameters: An elliptic-curve $E$ with order $N$ and a base point $G$. 
	\item[2.] The router $\mathcal{R}$ requests a remote attestation and public key from the validator $\mathcal{V}$ to register.
	\item[3.] Validator $\mathcal{V}$ calculates $\mathcal{A} = s\cdot G$, then sends $(\mathcal{A}, \sigma_{\mathcal{V}})$ to $\mathcal{R}$, where the $\sigma_{\mathcal{V}}$ is the remote attestation of validator $\mathcal{V}$'s encalve.
	\item[4.] $\mathcal{R}$ verifies  $\sigma_{\mathcal{V}}$. If the check passes, $\mathcal{R}$ generates a 256 bits random number $b$ and calculate $\mathcal{B} = b\cdot G$. Then send $\mathcal{B}$ to validator $\mathcal{V}$.
	\item[5.] $\mathcal{R}$ computes $k=b\cdot \mathcal{A}$, $\mathcal{V}$ computes $k = s\cdot \mathcal{B}$. Now $\mathcal{R}$ and $\mathcal{V}$ share a common communication key $k$ because $b\cdot \mathcal{A} = b \cdot s \cdot G = s \cdot b \cdot G = s \cdot \mathcal{B}$.
	\item[6.] It is difficult to calculate the value of $b$ from $\mathcal{B}$, so $\mathcal{B}$ can be propagated in untrusted channels. Other relay chain nodes can obtain the value $\mathcal{B}$ from validator $\mathcal{V}$. 
\end{itemize}

The above calculation can only be performed in the enclave, the communication key will be encrypted by the sealed key of the relay chain node's enclave and then stored on the disk.

The pseudo-code given in algorithm \ref{algorithm1} describes the process of secret sharing among n relay chain nodes, and the pseudo-code given in algorithm \ref{algorithm2} describes the entire process of communication key generation.

\subsection{Work Flow}
The workflow of our scheme includes two phases: key generation and registration phase and cross-chain session phase.

\subsubsection{Key Generation And Registration Phase} 
In this phase, the following activities will be performed:
\begin{itemize}
	\item Relay chain nodes initialize
	\item Key exchange between router and relay chain nodes.
	\item Registraction: Router shares the access control list with relay chain nodes.
\end{itemize}

\begin{figure}[h]
	\centering 
	\includegraphics[width=0.48\textwidth]{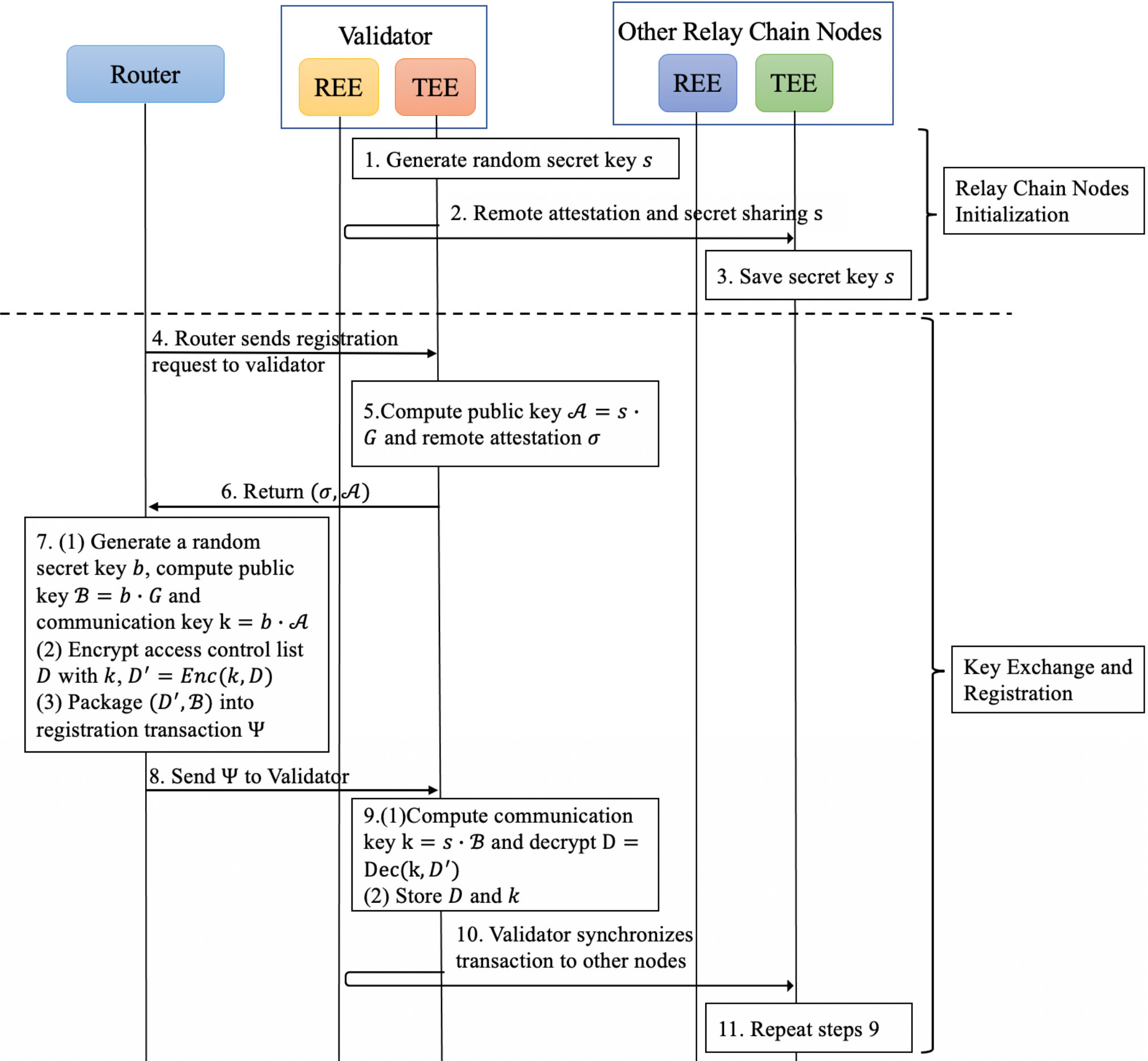} 
	\caption{Key generation and registration workflow}
	\label{workflow1}
\end{figure}

Fig.~\ref{workflow1} depicts the work flow of key generation and registration of parachain.

\subsubsection{Cross-chain Session Phase}
A cross-chain session is continuous communication in a short period of time for a certain business. The entire session starts with a cross-chain data request transaction, and ends with a cross-chain data response transaction with an empty Callback field.

\begin{figure}
	\centering 
	\includegraphics[width=0.48\textwidth]{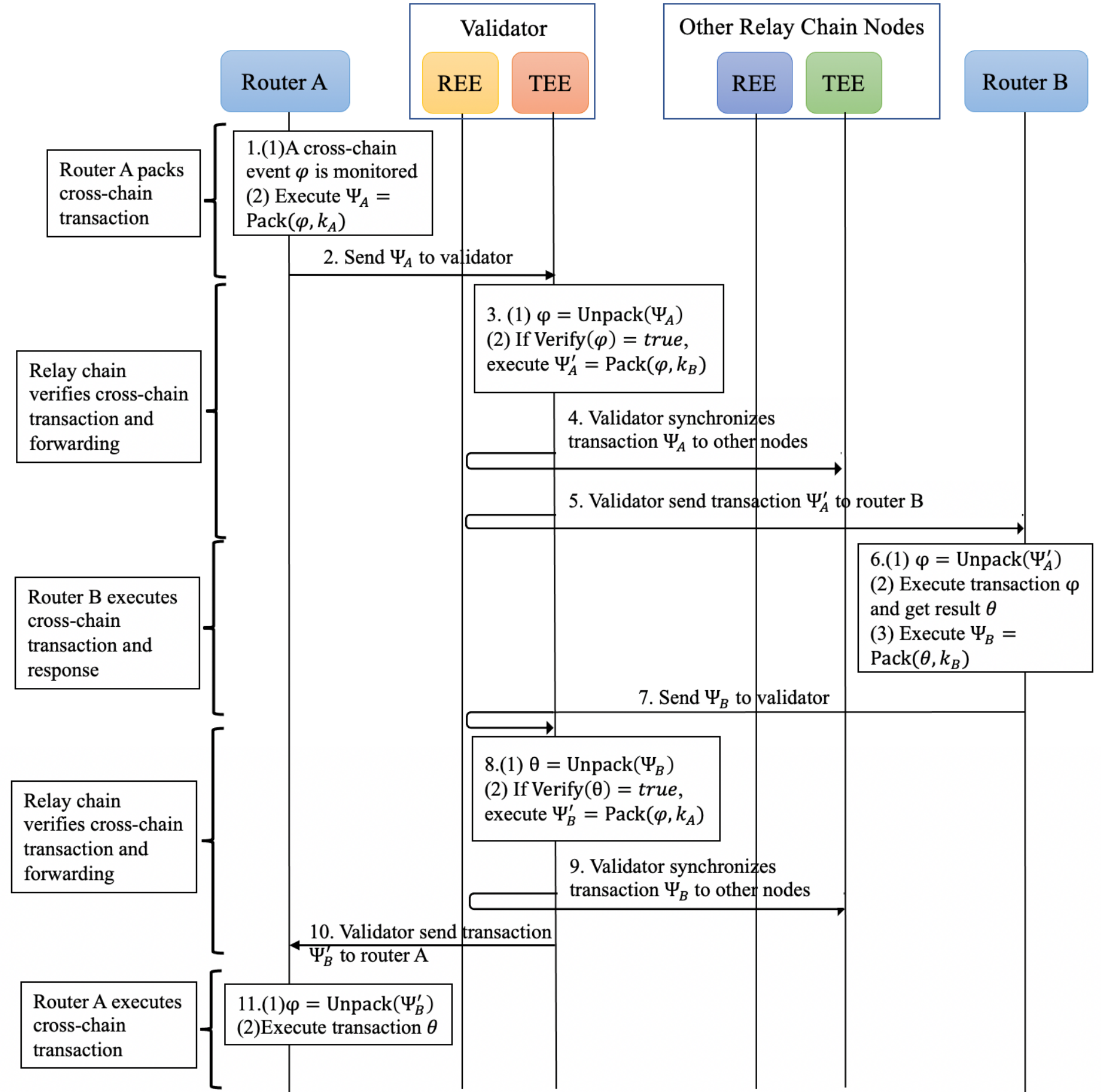} 
	\caption{Cross-chain Session Phase workflow}
	\label{workflow2}
\end{figure}

Fig.~\ref{workflow2} describes the shortest cross-chain session, that is, a cross-chain data request transaction and a cross-chain data response transaction. The operations involved in Fig.~\ref{workflow2} are:
\begin{itemize}
	\item $Pack(\varphi,k)$: the input $\varphi$ is a cross-chain transaction event which contains  all the information needed for the cross-chain format in Fig.~\ref{cross-chain-format}, and $k$ is the communication key. This function constructs a cross-chain transaction based on these two inputs and the protocol in Section~\ref{protocol}. 
	
	\item $Unpack(\Psi)$: the input $\Psi$ is a cross-chain transaction. This function first checks the signature of the router, and then decrypts the payload with the stored communication key. Function returns the decrypted payload.
	
	\item $Verify(\varphi)$: the input $\varphi$ is the payload referred to Fig.~\ref{cross-chain-format}. This function first checks the signature of the originator, and then verifies the access permission of this transaction. If all checks pass, return true, else false.
\end{itemize}

\section{Security Analysis}
\label{analysis}
Our sheme aims to ensure the privacy of cross-chain data. The combination of TEE and the relay chain reduces the attack surface, and it becomes difficult for an attacker to obtain data from the relay chain nodes. In this section we describe the main security properties provided by \tool.
\subsection{Data Confidentiality}
In order to ensure the confidentiality of cross-chain data, the communication key is used to encrypt the cross-chain data before it reaches the relay chain node. The decryption operation can only be performed in enclave to ensure that the cross-chain data is inaccessible to the outside world. Moreover, operations such as key generation, key computation, and secret sharing are also performed in the enclave environment to ensure that the key cannot be obtained by malicious attackers.  In addition, we use access control policies to ensure that unauthorized cross-chain operations cannot be executed, thus malicious attackers cannot obtain secret data through operations such as hash collisions and brute force enumeration.

\subsection{Resistance to Replay Attack}
Replay attack is one of the commonly used attack methods in the computer world. The attacker may obtain benefits by by replaying previous cross-chain transactions. In our solution, the transaction sequence number and timestamp can be used to defend against replay attacks.

\subsection{Resistance to Data Race}
In some cases, data race may occur. For example, two different users on a parachain request the same resources in a short period of time, and the cross-chain router preferentially processes the subsequent cross-chain transaction for benefit. But in our scheme, the use of transaction sequence numbers, timestamp, and session hash can prevent this from happening. The transaction sequence number and timestamp indicate the order of transaction processing, and the session hash can indicate which transaction to respond to.

\subsection{Resistance to Tampering and Masquerading Attack}
Assuming that a malicious attacker wants to tamper or Masquerade a cross-chain transaction, this behavior will be easily detected because of the signature. 

\subsection{Resistance to MITM Attack }
MITM (man-in-the-middle) attack is a cyberattack where the attacker secretly relays and possibly alters the communications between two parties who believe that they are directly communicating with each other. In our scheme, in the key exchange phase, the relay chain node can confirm the identity of the cross-chain router by verifying its certificate (public key) and signature, and the cross-chain router can verify the identity of the relay chain node through remote attestation. Therefore, it can defend against MITM attacks.

\subsection{Tracing Accountability}
In our scheme, an important role of the relay chain is to retain evidence in order to trace accountability. Consider a scenario where the router of the destination chain returns an execution result but the cross-chain transaction is not actually confirmed on the destination chain. Due to the loss caused to the user issuing the cross-chain transaction, the user will request to trace accountability, and the evidence will be found on the relay chain. The cross-chain router cannot deny it and will be punished.

\section{Performance Evaluation And Discussion}
\label{evaluation}
We have implemented \tool using Rust SGX SDK v1.1.3 for Linux and set up 3 relay chain nodes for our experimental evaluation. 
Relay chain nodes are with Intel(R) Xeon(R) CPU E3-1225 v5 and Ubuntu 18.04 operation systems. 
The parachains used for the experiment are Ethereum private chain and Quorum, and we respectively released a Hub contract on these two parachains specifically for publishing cross-chain events. 
The Ethereum private chain and its router run on MacOS laptop with Intel Core i7 CPU. 
Quorum and its router run on the same types of environment as relay chain nodes.
 
\begin{figure}[h]
	\centering 
	\includegraphics[width=0.45\textwidth]{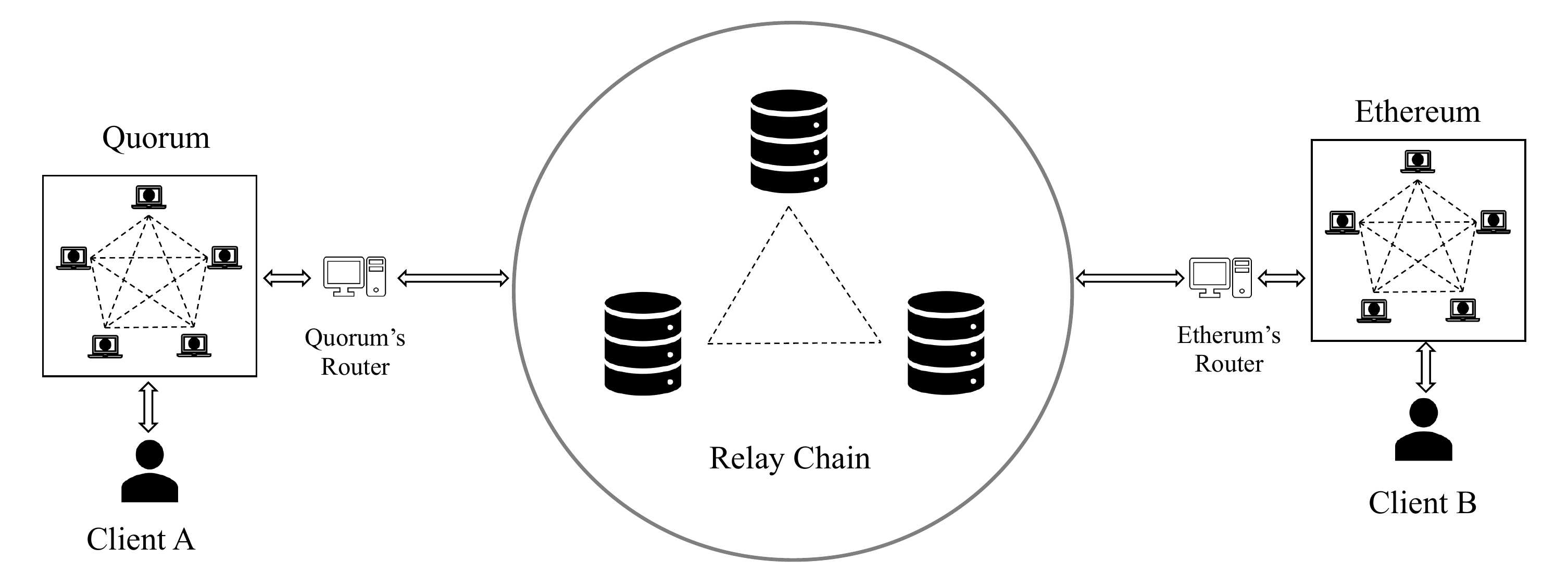} 
	\caption{Network structure of the experiment }
	\label{network}
\end{figure}

The network structure of the experiment was as Fig.~\ref{network}.

The first experiment tested the delay time of the complete execution of a cross-chain transaction, that is, the process from the client A sending a cross-chain request to the Ethereum router receiving and executing the transaction. The cross-chain transaction is a contract data read operation. The second experiment tested the additional time overhead of SGX for processing cross-chain transactions. The last experiment tested the memory overhead of SGX enclave.

\subsection{Latency of A Cross-chain Transaction}
\begin{figure}[h]
	\centering 
	\includegraphics[width=0.45\textwidth]{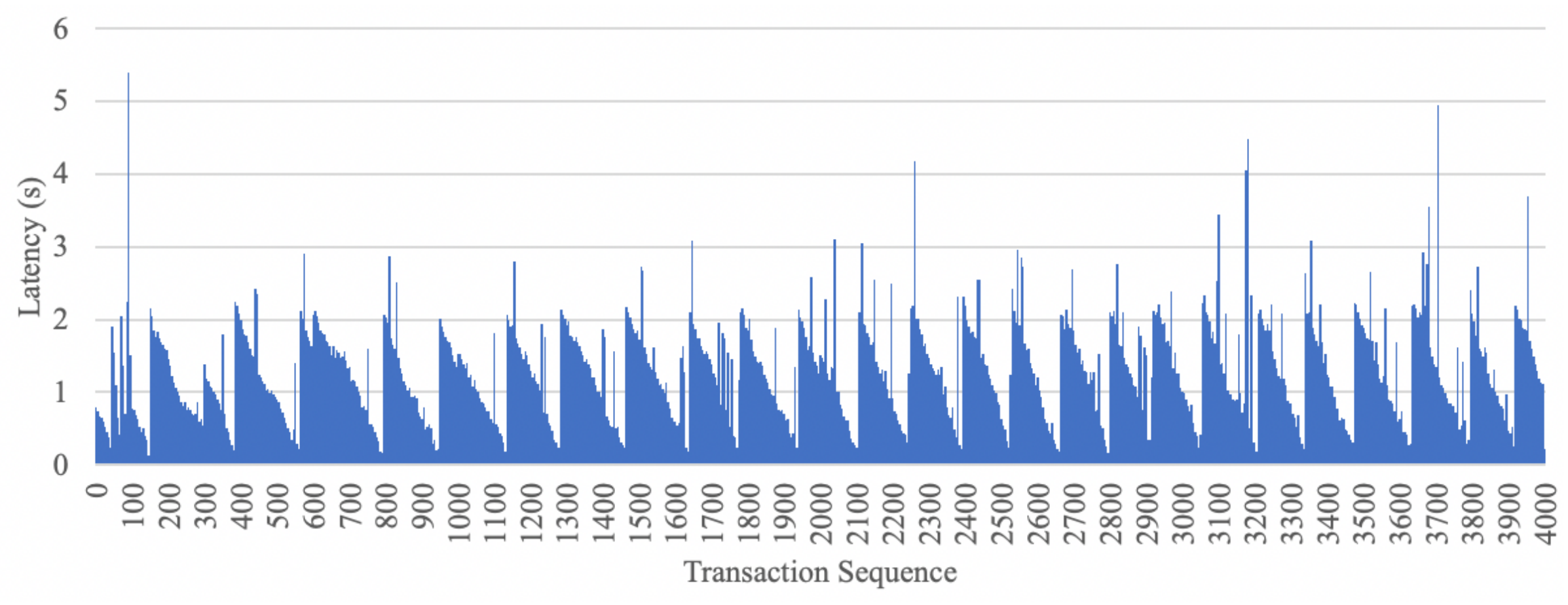} 
	\caption{Lantency of cross-chain transaction }
	\label{latency}
\end{figure}
In order to test the performance of our cross-chain system, we constructed a cross-chain event from Quorum Hub contract that requested data from an Ethereum smart contract. The whole process includes the following steps: Client A calls the Quorum Hub contract, Quorum confirms the transaction and issues the event, Quorum's router listens to the event and packages the cross-chain transaction, the relay chain verifies the cross-chain transaction, and the Ethereum's router receives and executes the transaction.

We tested 4000 cross-chain transactions. The results in Fig.~\ref{latency} show that the latency of a cross-chain transaction in our system is generally within 2s. We noticed that the time result graph shows a regular jagged shape, which may be due to Quorum packing transactions in batches in one block.

\subsection{The Impact of SGX Overhead on Performance}
\begin{figure}[h]
	\centering 
	\includegraphics[width=0.48\textwidth]{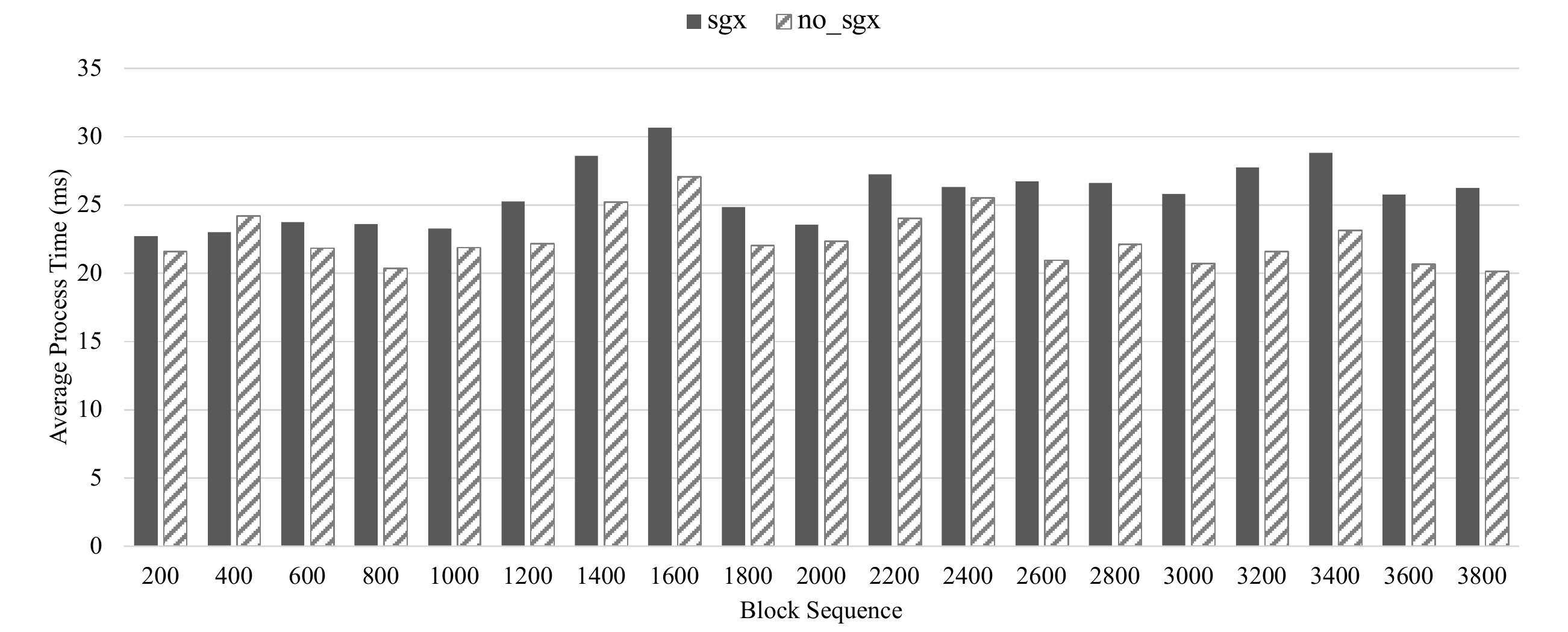} 
	\caption{Comparison of the processing time of relay chain nodes using SGX and not using SGX}
	\label{sgx_vs_no_sgx}
\end{figure}
Because we have used SGX to enhance the confidentiality of cross-chain transactions, some people may be suspicious of the additional overhead this will bring. Therefore, we designed an experiment to visually evaluate the performance impact of SGX.

According to Fig.~\ref{sgx_vs_no_sgx}, we have observed that the use of SGX by relay chain nodes does bring some overhead, but the impact of this overhead is relatively small. The average time overhead of using SGX to process a block on the relay chain is 25.8113ms, while the average time overhead of processing a block without SGX is 22.491628ms, therefore the additional cost of SGX is about 3.3ms. We consider that the performance loss caused by using SGX is completely acceptable, compared to the enhanced confidentiality that it brings.

\subsection{Memory Overhead of SGX Enclave}
\begin{figure}[h]
	\centering 
	\includegraphics[width=0.48\textwidth]{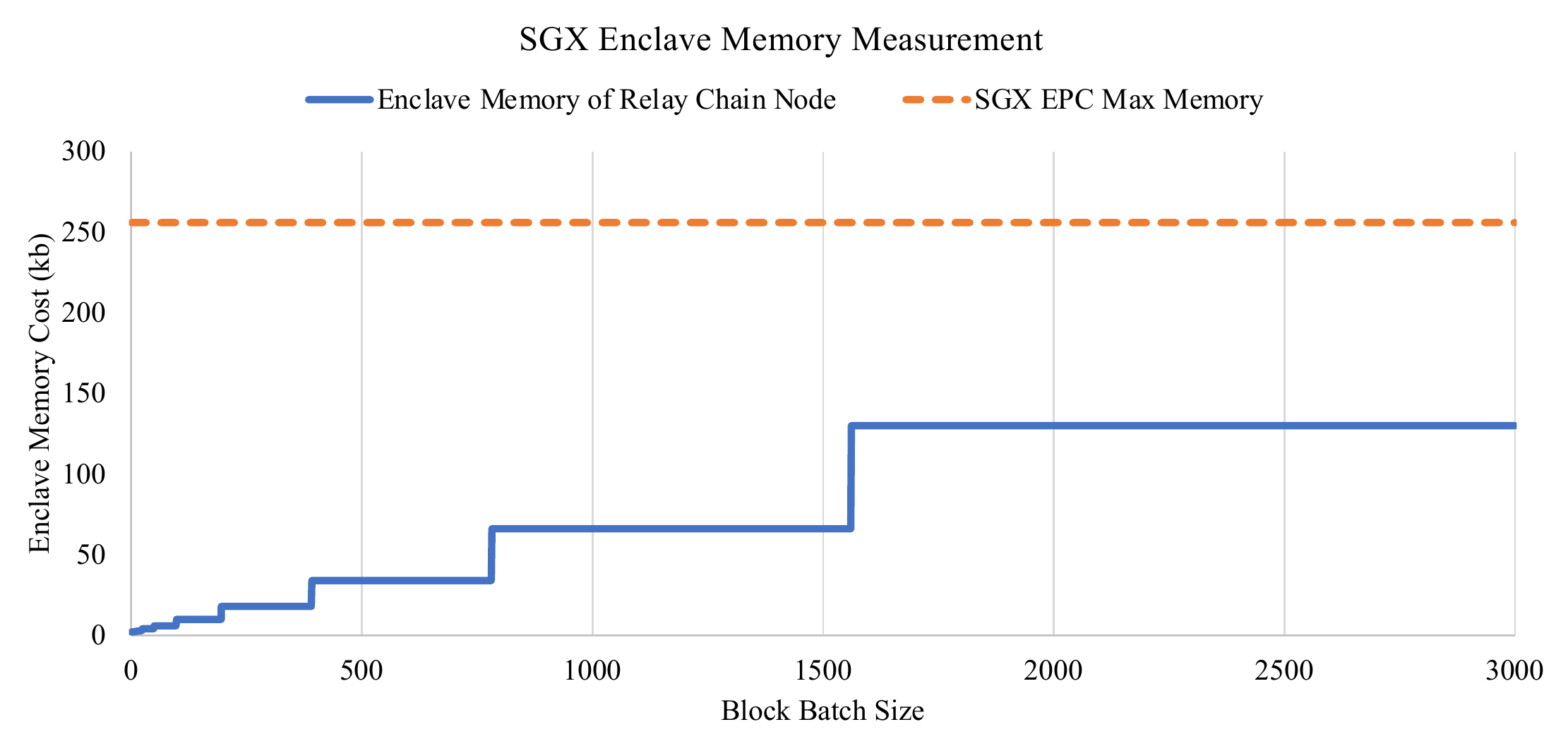} 
	\caption{Memory overhead of SGX enclave}
	\label{memory}
\end{figure}

In practical applications, SGX's EPC memory (maximum 256MB) limit will also cause concerns. Although there are some solutions to expand EPC memory, such as Aliyun ECS\cite{ali}, the mainstream SGX-supporting chips on the market still have a 256M memory limit.

For the consideration of SGX memory, we tested the changes in memory overhead as the batch size of the block's one-time packaged transactions increases.

The result is shown in Fig.~\ref{memory}. We observed that when the number of batches reaches 3000, the memory still has not reached the upper limit, which is sufficient to meet our cross-chain requirements. Therefore, the limitation of SGX memory will not cause the performance bottleneck of our cross-chain system, which means that the scalability of our system can be very good.

In addition, we can observe from the figure that the memory allocation mechanism of SGX enclave does not seem to be allocated on demand, but grows exponentially at 2MB, such as 1MB, 2MB, 4MB, 8MB, 16MB, 32MB, 64MB and so on. This means that when the memory allocation reaches the upper limit, the memory overhead for processing the block does not actually reach this value. Therefore, if we allocate SGX enclave memory according to actual needs, we can even extend the maximum block batch size to nearly twice the current one, which can be inferred from the figure.

\section{Related Work}
Multiple cross-chain schemes have been proposed. These schemes involve different technologies, such as notary scheme, hash-locking, relay, and side chain etc.

Ripple introduced an Interledger protocol (ILP) \cite{thomas2015protocol} for two different ledgers to convert currencies to each other through untrusted connectors. It uses ledger-provided escrow, that is conditional locking of funds, to allow secure payments through untrusted connectors.  The core idea of the ILP is that ledger-provided escrow guarantees the sender that their funds will only be transferred to the connector once the ledger receives proof that the recipient has been paid. Escrow also assures the connector that they will receive the sender’s funds once they complete their end of the agreement.

Back et al. \cite{back2014enabling} first introduced the concept of sidechain. In \cite{back2014enabling}, Back et al. proposed a technology pegged sidechains to enable Bitcoin and other ledger assets to be transferred between multiple blockchains. It allows to to transfer assets by providing proofs of possession in the transferring transactions themselves, avoiding the need for nodes to track the sending chain. 

Polkadot project\cite{wood2016polkadot} uses relay chain technology to solve cross-chain interaction needs. In Polkadot, there are the following three types of components: relay chain, parachain, bridge. Relay chain, the core of Polkadot, which coordinates the consensus and transactions between different parachains. Parachain collects and processes transactions. Bridge connects other heterogeneous blockchain such as Ethernet and BitNet. And there are four types of roles in Polkadot network: validators (responsible for verifying the data of parachains), collectors (responsible for collecting data from parachains and submitting them to the validators), nominators (providing deposits and credit endorsements for validators), and phishing Person (responsible for reporting and proving malicious behavior).

Cosmos\cite{cosmos2021} is a network connecting many independent blockchains, called zones. The first zone on Cosmos is called the Cosmos Hub, which serves as a relay chain. The hub and zones of the Cosmos network communicate with each other via an inter-blockchain communication (IBC) protocol, a kind of virtual UDP or TCP for blockchains. The zones are powered by Tendermint Core, which provides a high-performance, consistent, secure PBFT-like consensus engine.

Relay chain technology is a very popular cross-chain solution in practical projects. In addition to Polkadot and Cosmos mentioned in the previous article, there are also BitXhub\cite{BitXHub}, WeCross\cite{WeCross}, Hyperledger Cactus\cite{cactus} and other solutions that also use relay chains.
\label{relatedwork}

\section{Conclusion And Future Work}
Data privacy in cross-chain interoperability is still a challenge. In this paper, we proposed \tool, a novel cross-chain architecture that can enhance the confidentiality of cross-chain data with TEE. TrustCross runs a cross-chain interoperability protocol to unify the standards of cross-chain transactions to transfer messages between different blockchains. And the fine-grained access control mechanism in our proposal can prevent malicious attackers from accessing unauthorized resources through guessing and brute force enumeration. In addition, to avoid the leakage of cross-chain data during transmission, we designed a secure key exchange algorithm to generate a communication key for data encryption, and this algorithm also ensures that the key will not be leaked. 
The confidentiality guarantee of TEE, combined with the encryption of cross-chain data and the fine-grained access control mechanism are the key points to ensure privacy in our proposal.
The experimental data shows that \tool has a reasonable latency and high scalability on executing cross-chain transactions across heterogeneous blockchains.

Our work is based on the security guarantee of TEE, but we have noticed that there are some attacks against TEE, such as side channel attacks. For future work, we hope to focus on reducing some of the harm caused by side-channel attacks. In addition, proof of the validity of cross-chain information and proof of execution results will also be our future research focus.
\label{conclusion}


\bibliographystyle{IEEEtran}
\bibliography{trustcross}

\end{document}